# Inflation and unemployment in Japan: from 1980 to 2050

Ivan O. Kitov
Institute for the Geospheres Dynamics, Russian Academy of Sciences
Russia, Moscow, Leninsky prospect, 38/1
e-mail: ikitov@mail.ru
mobile: +43-0664-3919022

Abstract

The evolution of inflation, $\pi(t)$, and unemployment, $UE(t)$, in Japan has been modeled. Both variables were represented as linear functions of the change rate of labor force, $dLF/LF$. These models provide an accurate description of disinflation in the 1990s and a deflationary period in the 2000s.

In Japan, there exists a statistically reliable ($R^2$=0.68) Phillips curve, which is characterized by a negative relation between inflation and unemployment and their synchronous evolution: $UE(t) = -0.94\pi(t) + 0.045$. Effectively, growing unemployment has resulted in decreasing inflation since 1982.

A linear and lagged generalized relationship between inflation, unemployment and labor force has been also obtained for Japan: $\pi(t) = 2.8*dLF(t)/LF(t) + 0.9*UE(t) - 0.0392$.

Labor force projections allow a prediction of inflation and unemployment in Japan: CPI inflation will be negative (between -0.5% and -1% per year) during the next 40 years. Unemployment will increase from ~4.0% in 2010 to 5.3% in 2050.

## Introduction

In a series of research papers we have carried out a thorough investigation of the change rate of labor force level as the single driving force behind inflation and unemployment - Kitov (2006ab; 2007ab); Kitov, Kitov and Dolinskaya (2007ab). In this framework, inflation in Japan was successfully modeled for the period between 1981 and 2003 (Kitov, 2006c). It has been demonstrated in all previous studies conducted for the USA, Japan, France, Austria, Germany and Canada that there exists a linear and potentially lagged link between labor force, inflation and unemployment. In some countries, this generalized relation can be split into two independent linear links between inflation and labor force and between unemployment and labor force. Obviously, these individual linear dependencies on the same defining variable result in the existence of reliable Phillips curves in these countries. These Phillips curves are not defined in standard form, however, since in many cases they are represented by lagged dependences between inflation and unemployment, any of these variables likely to be in. An important feature of our empirical study consists in the fact that coefficients in these linear dependencies sometimes are positive and sometimes are negative. In the former case (positive slope), increasing inflation is associated with increasing (but lagged) unemployment, as is it observed in the USA (Kitov, 2006ab). In the latter case, increasing inflation results in a decreasing unemployment rate, as observed in Germany (Kitov, 2007b).

This paper is primarily aimed at the estimation of the generalized relationship between the change rate of labor force level, $LF(t)$, inflation, $\pi(t)$, and unemployment, $UE(t)$, in Japan. Also, individual relationships between the change rate of $LF$ and inflation, and the change rate of $LF$ and unemployment are investigated. Such relationships allow answering important practical questions addressed in numerous studies of inflation and unemployment. For example: Do central banks really affect inflation and unemployment when conducting monetary policy?

Leigh (2004) studied influence of monetary policy on the liquidity trap in Japan. In other words, was there some monetary policy which the Bank of Japan could conduct to avoiding deflationary slump? He found that the trap arose not because of monetary policy mistakes and that “a policy of responding more aggressively to the inflation gap while keeping the low inflation target would have provided little improvement in economic performance”. This conclusion is partly in line with our findings, which deny any possibility of the influence on inflation except that transmitted through inflation dependence on labor force and unemployment.

There exists, however, a common opinion in the economists and central bankers community that inflation is a monetary phenomenon. Nelson (2006) investigated this assumption as applied to Germany and Japan and argued that the experiences of these countries in the 1970s indicate that once inflation is accepted by policymakers as a monetary phenomenon, the main obstacle to price stability has been overcome. Hence, central banks are able to control inflation through monetary policy.

Currently, many popular inflation models are concentrated around the New Keynesian Phillips Curve (NKPC) approach. De Veirman (2007) studied, in the NKPC framework, the output-inflation trade-off in Japan as a linear relationship with a time-invariant slope during the period between 1998 and 2002. He found that large negative output gap did not cause accelerating deflation, which would be expected according to the NKPC. Kamada (2004) investigated the importance of various real-time measures of output gap for inflation prediction and development of monetary policy by the Bank of Japan and reported that some measures of output gap to be marginally useful for the inflation prediction despite problems with high uncertainty in real-time estimates. The Taylor rule needed more ingredients "for preventing for preventing the asset bubble". These findings do not contradict our model since deflation is a natural result of decreasing labor force level in Japan, not output gap.

Sekine (2001) studied inflation function and forecasts at one-year horizon for Japan using equilibrium correction model. He demonstrated only marginal forecast improvement, relative to the simplest AR model, even when such variables as markup relationships, excess money and output gap are included. Feyzioğlu and Willard (2006) found that foreign countries, specifically China, have no influence on prices in Japan. These results also confirm the dependence of inflation only on labor force.

The evolution of unemployment in Japan was also thoroughly studied. Caporale and Gil-Alana (2006) tested unemployment time series in Japan for structural breaks at unknown points. They showed that "structuralist" approach to unemployment works well in Japan and interpreted this observation using specific features of labor market. Only one structural break was identified in the Japanese unemployment time series. Pascalau (2007) found a long-run equilibrium relation between unemployment rate, productivity, and real wages in Japan. All the involved variables had a unit root and, thus, cointegration tests with non-linear error-correction mechanisms were applied. He reported relatively longer persistency of unemployment shocks. In our framework, these findings are random and reflect statistical properties of the evolution of labor force level in Japan.

Kitov (2006c) estimated for Japan empirical coefficients in the inflation model based on the link between inflation and labor force change. A standard best-fit procedure applied to cumulative values of CPI inflation (with imputed rent) gave a slope of 1.77 and constant term -0.0035. It was also found that the change in labor force occurred practically simultaneously with that in inflation. This paper extends the period of the measured inflation modeling to 2006, and also provides a prediction of inflation to 2050. The evolution of unemployment is also modeled as a linear lagged function of labor force. From the inflation and unemployment models, a Phillips curve for Japan is estimated. Finally, a generalized relationship between inflation, unemployment and the change rate of labor force level is obtained.

The remainder of this paper consists of two Sections and Conclusion. Section 1 is devoted to the constriction of a Phillips curve for Japan. Section 2 presents quantitative results for the generalized link between three involved variables. Labor force projections are used to predict the evolution of inflation and unemployment between 2007 and 2050. In Conclusion, some principal findings are highlighted.

## 1. The Japanese Phillips curve

Data on labor force, inflation, and unemployment were obtained from various sources. The Statistics Bureau (SB, 2006) of the Ministry of Internal Affairs and Communications provides information on various economic and demographic variables. The U.S. Bureau of Labor Statistics (BLS, 2007) provides two sets of data: one obtained according to national definition (NAC) and another obtained according to US definition of corresponding variable. Japan is a country with a modern statistical service. The Organization of Economic Cooperation and Development (OECD, 2006) along with Eurostat (ES, 2006) publishes very useful data on labor force measurements for a relatively long period.

There are several measures for inflation. The most popular definitions for the overall price change are GDP deflator and Consumer Price Index. In many countries, the CPI definition was extended recently by inclusion of imputed rent. Thus, various inflation time series might be studied, but only two of them are used in this paper. Figure 1 shows corresponding curves for these two inflation estimates: the OECD GDP deflator and CPI provided by the Japanese Statistical Bureau. The difference between the curves is minor but very illustrative. The GDP deflator curve is below that for the CPI inflation since 1990. One has to bear in mind that the latter variable is a constituent part of the former one. The accuracy of CPI inflation measurements in Japan is also under doubt as discussed by Shiratsuka (1999) and Ariga and Matsui (2002).

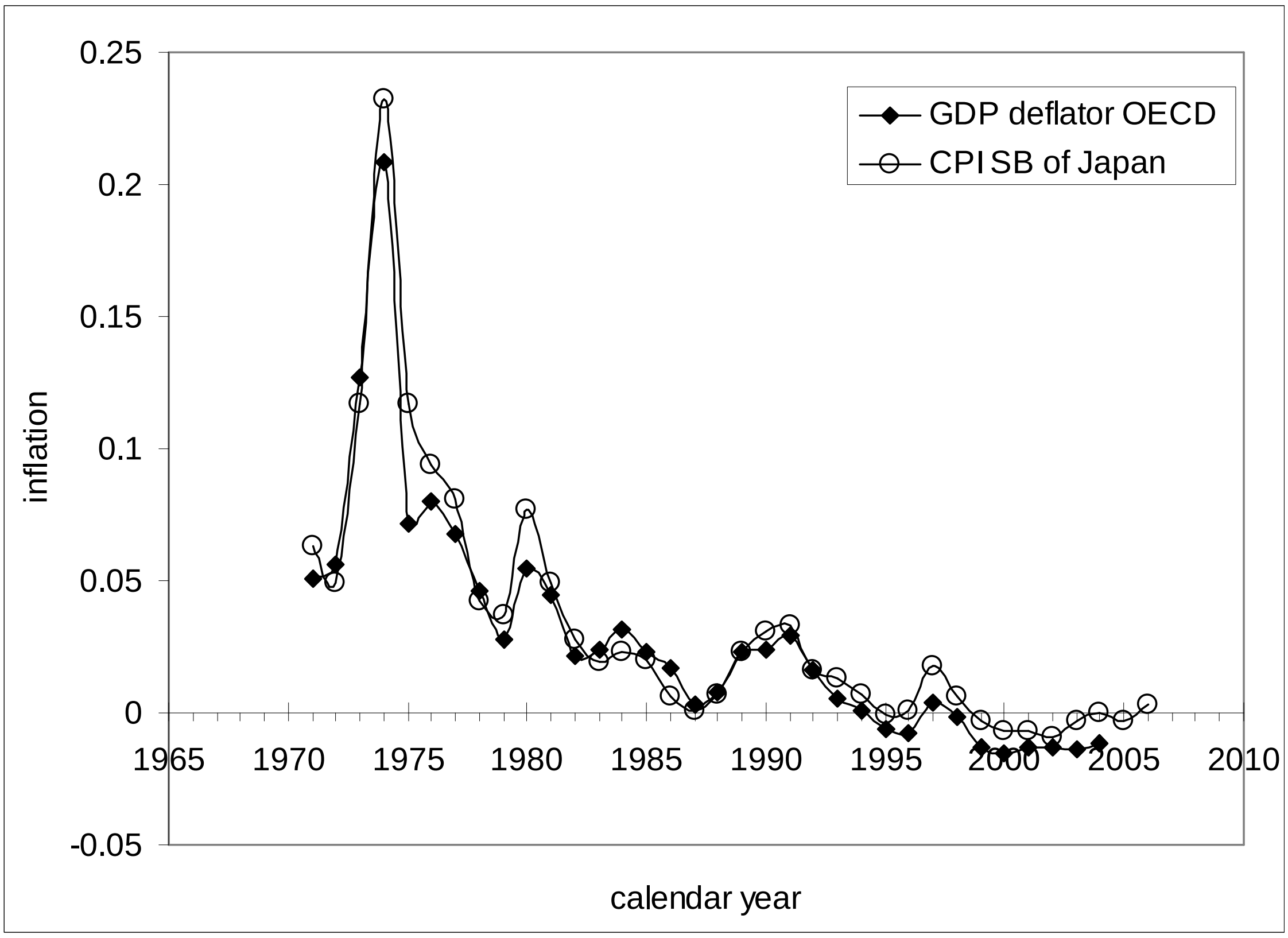

Fig. 1. Comparison of two measures of inflation: CPI and GDP deflator. The curves are slightly different during some periods of time. Notice that the GDP deflator curve is below that for CPI after 1990. The gap between the curves seems to be growing over time.

For any quantitative analysis, the most important issue is the quality of corresponding measurements. There are two main requirements to these data: they have to be as precise as possible in respect to any given definition, and the data must by comparable over time. The precision is related to methodology of measurements and implementation of corresponding procedures. The comparability is provided by the consistency of definitions and methodology. For example, the OECD (2005) provides the following information on the comparability of labor force and unemployment time series for Japan:

> ***Series breaks:*** *In 1967 the "household interview" method was replaced by the "filled-in-by-household" method and the survey questionnaire was revised accordingly.*

According to this statement one should not expect any breaks in linear relationship between the studied variables: labor force, inflation, and unemployment. However, as shown in below, the Phillips curve for Japan demonstrates a break, which indicates the presence of some other problems in the general comparability of the measurements before and after 1982.

We use two different estimates of unemployment in Japan provided by national statistics and according to US definition. Figure 2 demonstrates that they are very close and almost undistinguishable before 2000. Kitov (2006c) also used an unemployment series provided by the OECD. This series was also close in shape to those in Figure 2, but underwent a significant divergence after 1974. True unemployment, as related to some perfect (but not currently available) definition of unemployment, might be between these curves and out of the curves as well. At the same time, both presented measures of unemployment are similar and it is likely that the true unemployment accurately repeats their shape. In this case, any of the measures can be used in quantitative modeling as representing the same portion of the true unemployment. Similar statement is valid for inflation measures. Apparently, actual problems are associated not with the difference between measured and true variables but with sudden jumps in the definitions of measured variables.

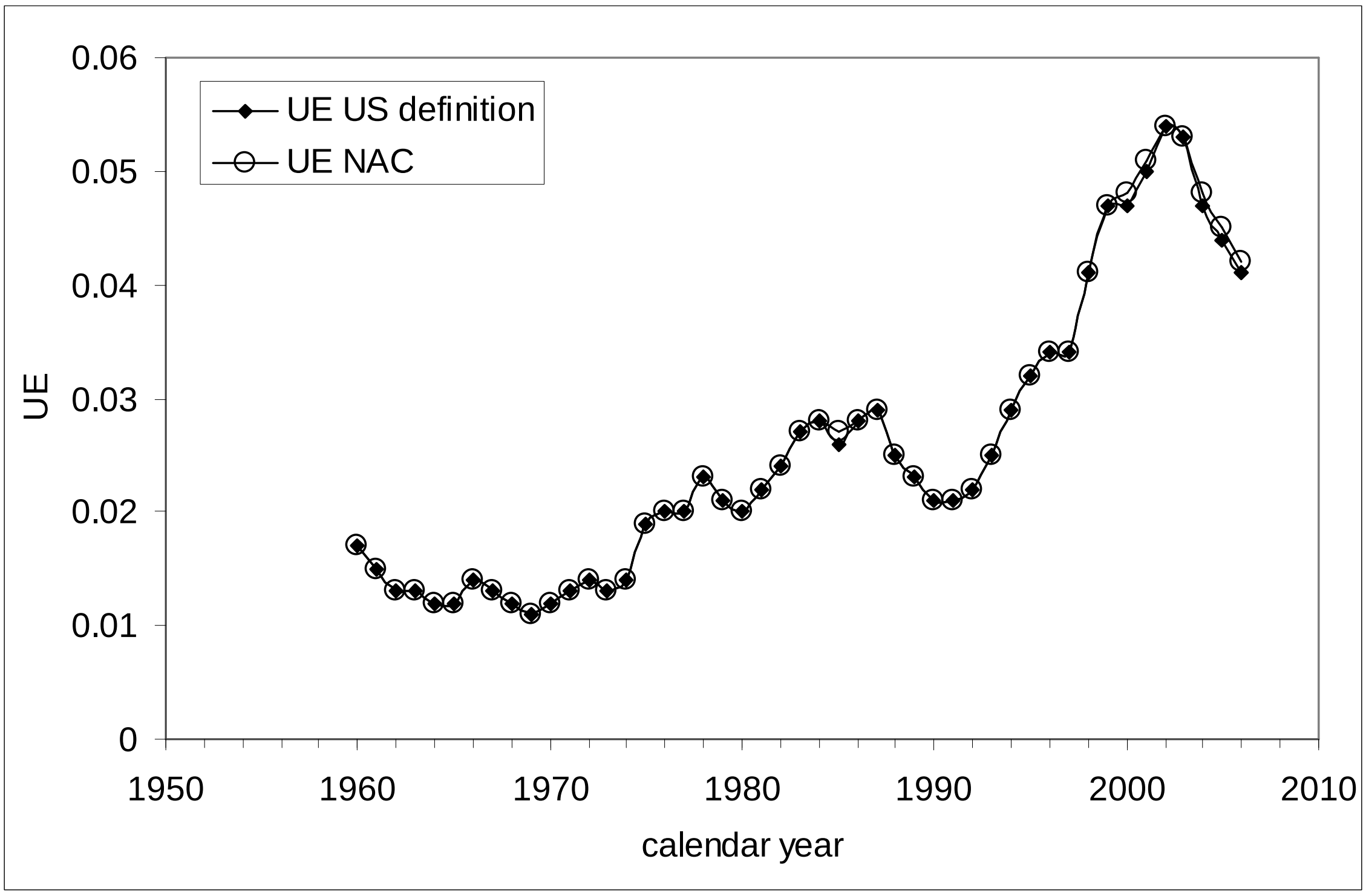

Figure 2. Comparison of two definitions of unemployment. The curves are slightly different since 2000.

The trade-off between the rate of change in money wage (later inflation) and unemployment was originally introduced by A.W. Phillips. Since 1958, the concept related to the Phillips curve has undergone numerous revisions. A good review of the current *status quo* is given by Rudd and Whelan (2005). As discussed above, we obtained different results when modeling the Phillips curve in developed countries. Only the same driving force behind inflation and unemployment unify them and allow building a Phillips-curve-type relation between them.

The links between inflation and unemployment actually demonstrate various and even opposite dependencies. In the USA, this dependence is characterized by a positive influence of inflation on unemployment (Kitov, 2006a). Effectively, low inflation in the USA leads low unemployment by three years because of empirically estimated 3-year lag between these two variables. Germany (Kitov 2007b) provides a case with a negative slope, i.e. low unemployment results in high inflation.

Figure 3 presents a scatter plot for unemployment (NAC) and CPI inflation in Japan. A linear regression gives a negative slope of -0.94 and constant term 0.041 for the period between 1982 and 2006. This regression has been calculated with various time shifts between the unemployment and inflation time series. The best fit ($R^2$=0.68) was obtained in the case when the unemployment curve and the inflation curve are not shifted. Before 1982, there is no linear relation between unemployment and inflation, as Figure 4 demonstrates, where the CPI inflation curve is modified according to the results of the linear regression in Figure 3.

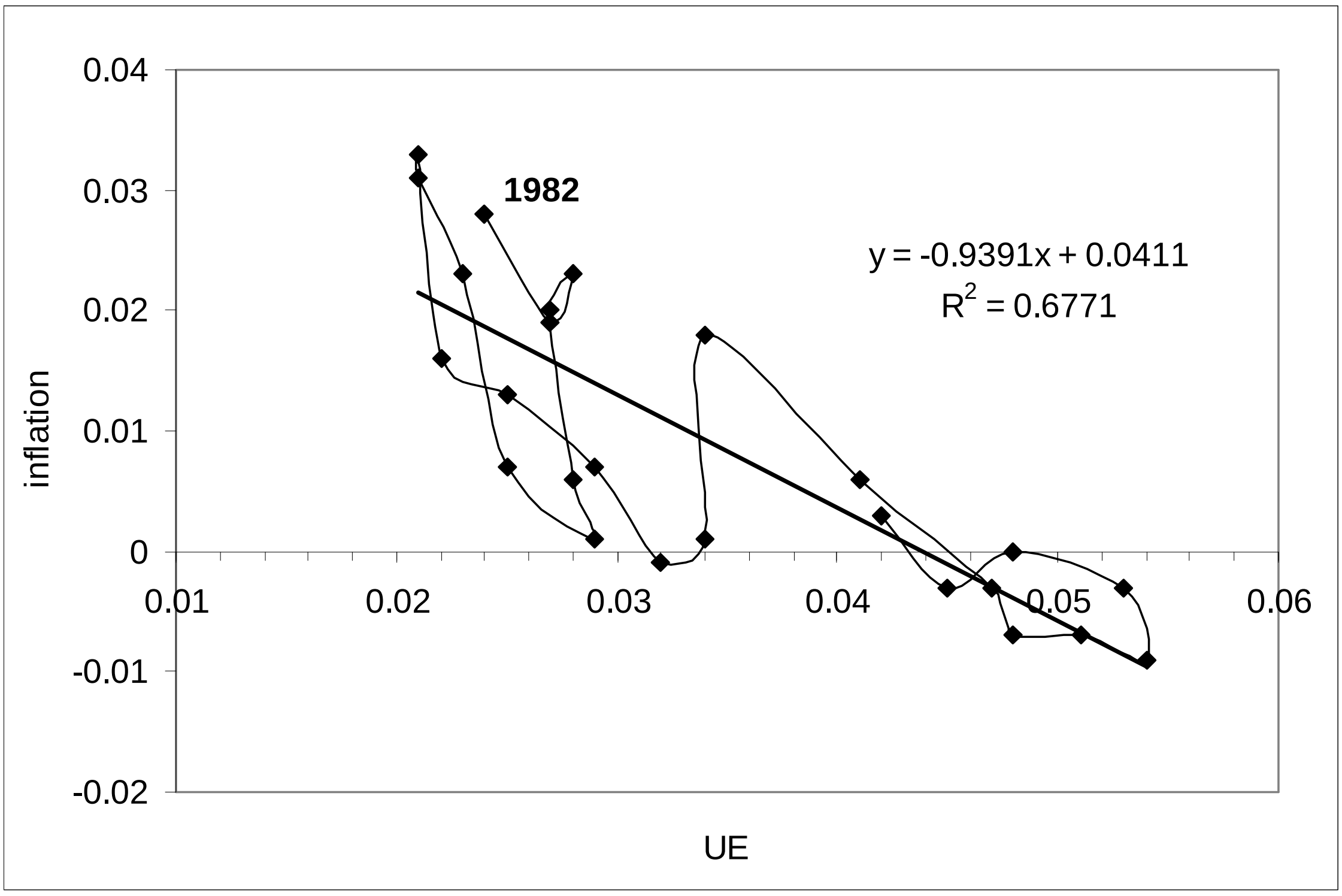

Figure 3. Inflation/unemployment scatter plot and linear regression for the period between 1982 and 2006. Neighboring years are connected by a curve. This curve represents the Phillips curve for Japan. $R^2$=0.68. Regression coefficients are -0.94 and 0.041. Therefore, increasing inflation leads to decreasing unemployment and vice versa. Similar link between inflation and unemployment is observed in Germany.

The slope obtained by linear regression is negative. The same situation is observed in Germany, where unemployment leads inflation by one year. These countries also have similarities in their inflation history (Nelson, 2006). In the USA, where inflation leads unemployment, the slope is positive. This swap of the lead is likely to be the reason for the difference in the sign of the slope in the Phillips curves for USA and Japan.

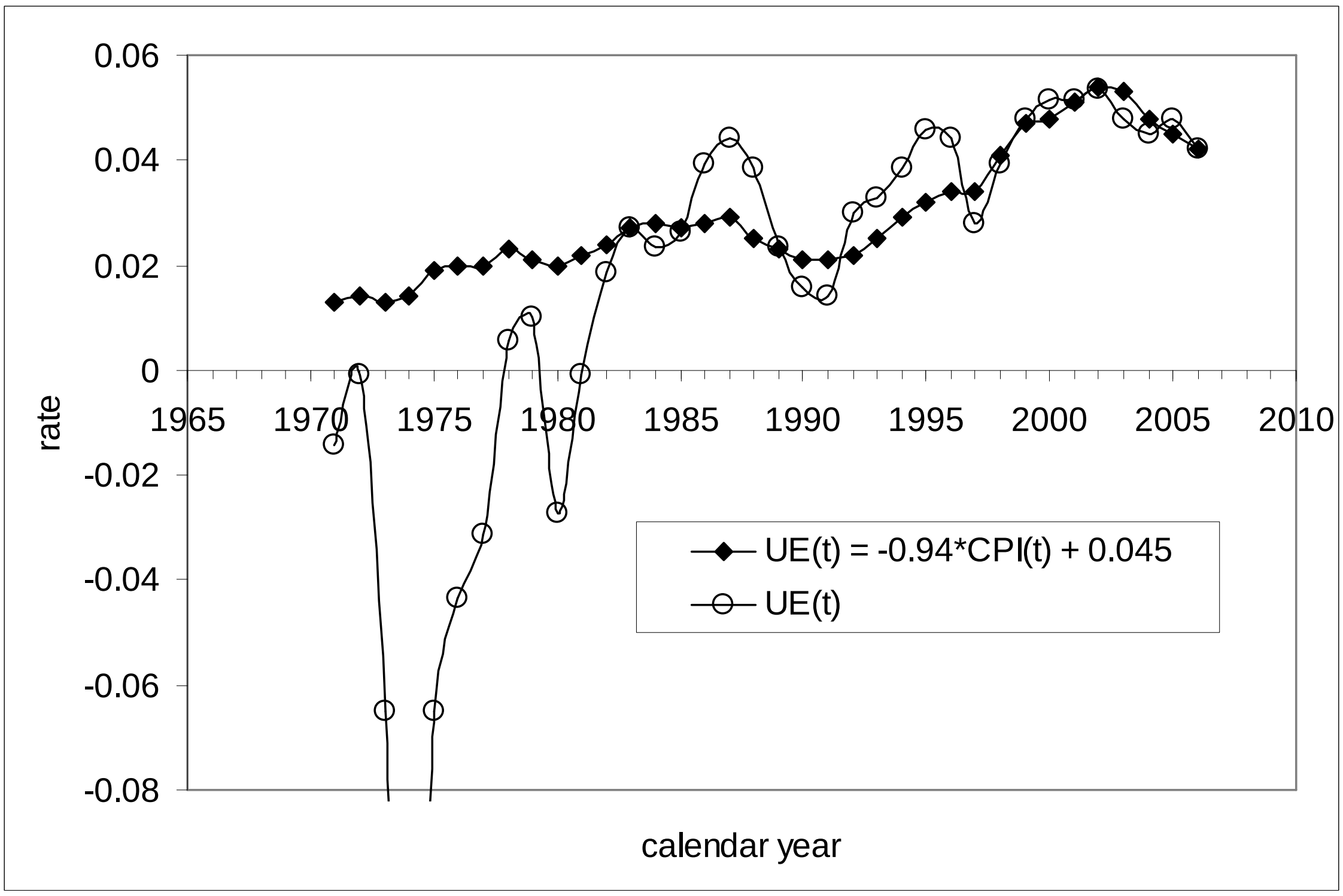

Figure 4. Measured unemployment and that predicted from CPI inflation for the period between 1971 and 2006 using the results of linear regression in Figure 3. Free term in the relationship is elevated by 0.004 in order to better describe the last ten years. Before 1982, the curves diverge.

Figure 4 demonstrates a relatively good agreement between the measured and predicted curves and Figure 3 actually provides the Japan's Phillips curve:

$$UE(t) = -0.94[0.14]\pi(t) + 0.045[0.005] \quad (1)$$

Standard deviation of the difference between the predicted and measured curve is *stdev=0.007*. Statistical estimates show a relatively high reliability of the Japan's Phillips curve, especially during the last 15 years. Therefore, one can expect that decreasing inflation in the years to come will be accompanied by increasing unemployment.

The existence of the Phillips curve in Japan raises a question about the consistency of monetary policy of the Bank of Japan. Does the bank conduct a monetary policy, which balances inflation and unemployment? Unlike Germany, where the Bundesbank has been showing during the last twenty five years the unwillingness to reduce unemployment in exchange for higher inflation, the BoJ was not able to decrease unemployment in order to get positive inflation figures.

## 2. Modeling inflation and unemployment in Japan

As many economic parameters, labor force estimates are also agency dependent due to various definitions and different population adjustments. Figure 5 compares the change rate of labor force provided by the OECD (2006), Eurostat (2006), according national and US definition (BLS, 2007). Despite strong similarity, some discrepancy reaching 0.1 (or 10% of the total labor force) is observed. Such a difference is an important indicator of the difficulties in labor force definition. Further investigations are necessary to elaborate a consistent understanding of the term "labor force". The model linking labor force change and inflation is likely a good candidate for quantitative consolidation of various definitions and approaches.

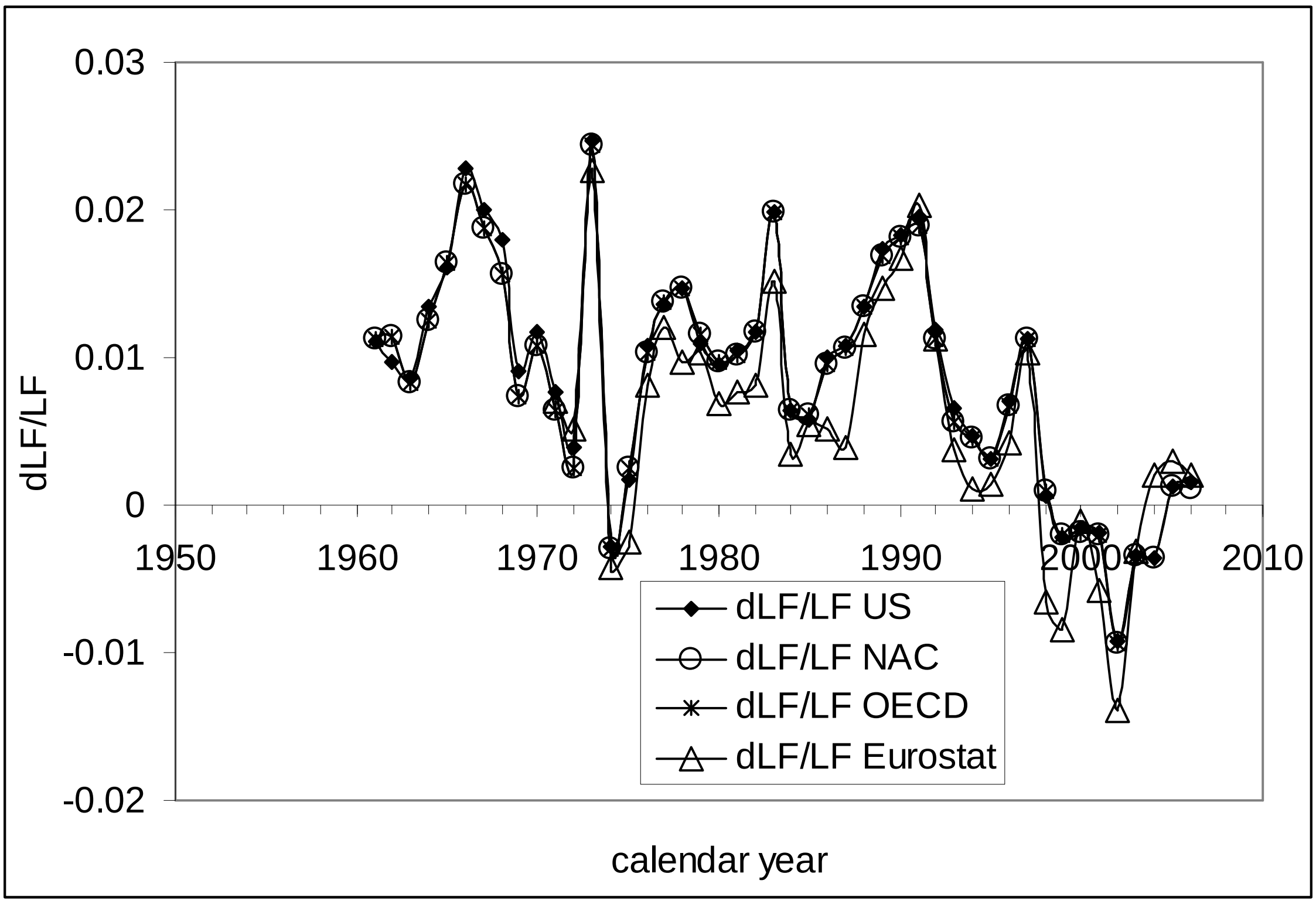

Figure 5. Comparison of four versions of the change rate of labor force level in Japan: Eurostat, OECD, national accounts (NAC) and US definition.

There exists a generalized relationship linking inflation and unemployment to the change rate of labor force level. Therefore we analyze CPI and GDP deflator in Japan in relation to unemployment and labor force level according to standard procedure described in previous papers. The preliminary step consists in an inspection of general features of all involved time series, as described above.

First, we test the existence of a link between inflation and labor force. Because of the structural break in the 1980s, we have chosen the period after 1982 for linear regression analysis. Varying time lag between labor force and inflation time series one can obtain the best-fit coefficients for the prediction of CPI inflation, $\pi(t)$, according to the relationship:

$$\pi(t) = A + BdLF(t-t_0)/LF(t-t_0) \quad (2)$$

where $A$ and $B$ are constants and $t_0$ is the time lag, which can be zero or some positive value. Figure 6a depicts the best-fit case with $A$= 0.0007 [0.002], $B$= 1.31 [0.19], and $t_0$=0 years. Kitov (2006c) obtained slightly different coefficients for the period between 1981 and 2003, but these differences are negligible. Because of the shortness of the modeled period, the estimate of coefficient $B$ is not very reliable. There is no time lag between inflation and unemployment in Japan. Coefficient *A*, defining the level of inflation in the absence of labor force change, practically is undistinguishable from zero.

A more precise and reliable method to compare observed and predicted inflation consists in comparison of cumulative curves (Kitov, Kitov, Dolinskaya, 2007ab). Short-term oscillations and uncorrelated noise in data induced by inaccurate measurements and a bias in definitions of the measured variables are smoothed out in cumulative curves. Any actual deviation between two cumulative curves persists in time if measured values are not matched by defining relationship. Predicted cumulative values are very sensitive to the coefficients in relationship (2). Therefore, we use in Figure 6a the coefficients obtained from the matching process between cumulative curves shown in Figure 6b. The cumulative curves are characterized by complex shapes. There are periods of intensive inflation growth and a deflationary period. The labor force change, defining the predicted inflation curve, follows all the turns in the measured cumulative inflation. One can conclude that relationship (2) is valid and the labor force change is the driving force of inflation.

For obvious reasons, it is difficult to precisely estimate the change in labor force level during one year. However, there are some benchmark years when all previous estimates are revised in order to match some better measured level of labor force. So, one can expect an increasing relative precision of the change in labor force level with increasing time baseline. The net change during 10 years should be measured with lower relative uncertainty than during one year.

Second step consists in the modeling of unemployment as a function of labor force change. Figure 7 presents the results of a simple manual trail-and-error matching process for the period between 1980 and 2006. Since such a procedure is based on visual fit only, no statistical estimates are made. The resulting relationship between unemployment and labor force in Japan is as follows:

$$UE(t) = -1.5*dLF(t)/LF(t) + 0.045 \quad (3)$$

The observed and predicted curves demonstrate similar shapes between 1980 and 2006 with higher volatility in the predicted unemployment. An important feature of (3) is the negative relation between unemployment and labor force. Any increase in the labor force level between 1980 and 2006 resulted in a decrease in unemployment. Such a trade-off provides a useful tool to treat high unemployment – one needs to increase labor force somehow.

The ultimate part of the modeling gathers three individual relationships in one generalized relation. So, we are trying to find the best-fit coefficients for the generalized equation:

$$\pi(t) = D_1*dLF(t)/LF(t) + D_2*UE(t) + D_3 \quad (4)$$

As before, the estimation of coefficient is (4) is based on the procedure developed in (Kitov, Kitov, Dolinskaya, 2007ab). The best-fit coefficients providing the lowermost RMS deviation between cumulative curves are as follows: $D_1$=2.8, $D_2$=0.9, $D_3$=-0.0392. Figure 8 depicts this case with the NAC definition of labor force. The evolution of the cumulative curves (Figure 8b) of the observed and predicted CPI inflation is very close. Therefore, the three involved variables are linked by an equilibrium long run relation.

a)

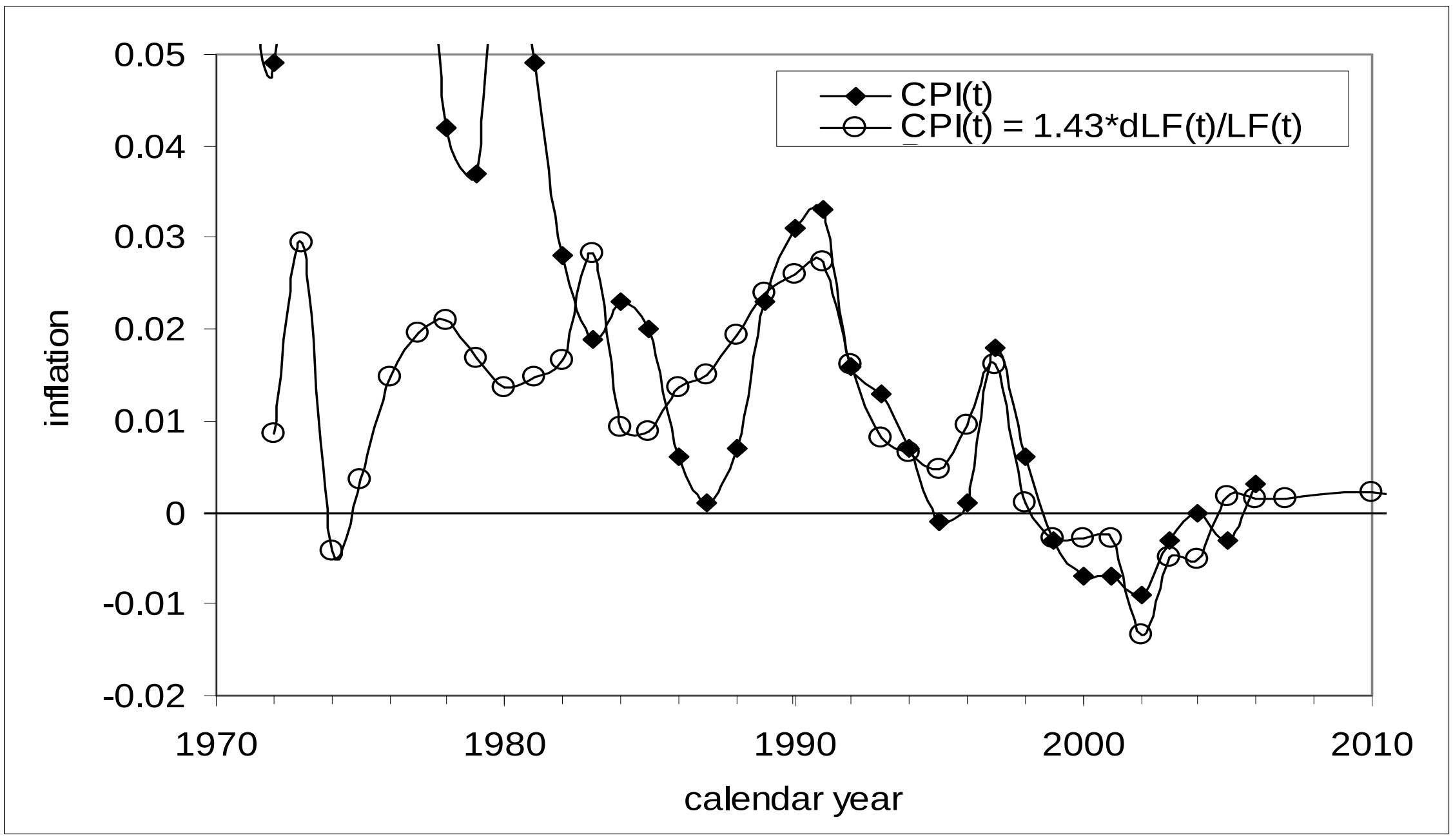

b)

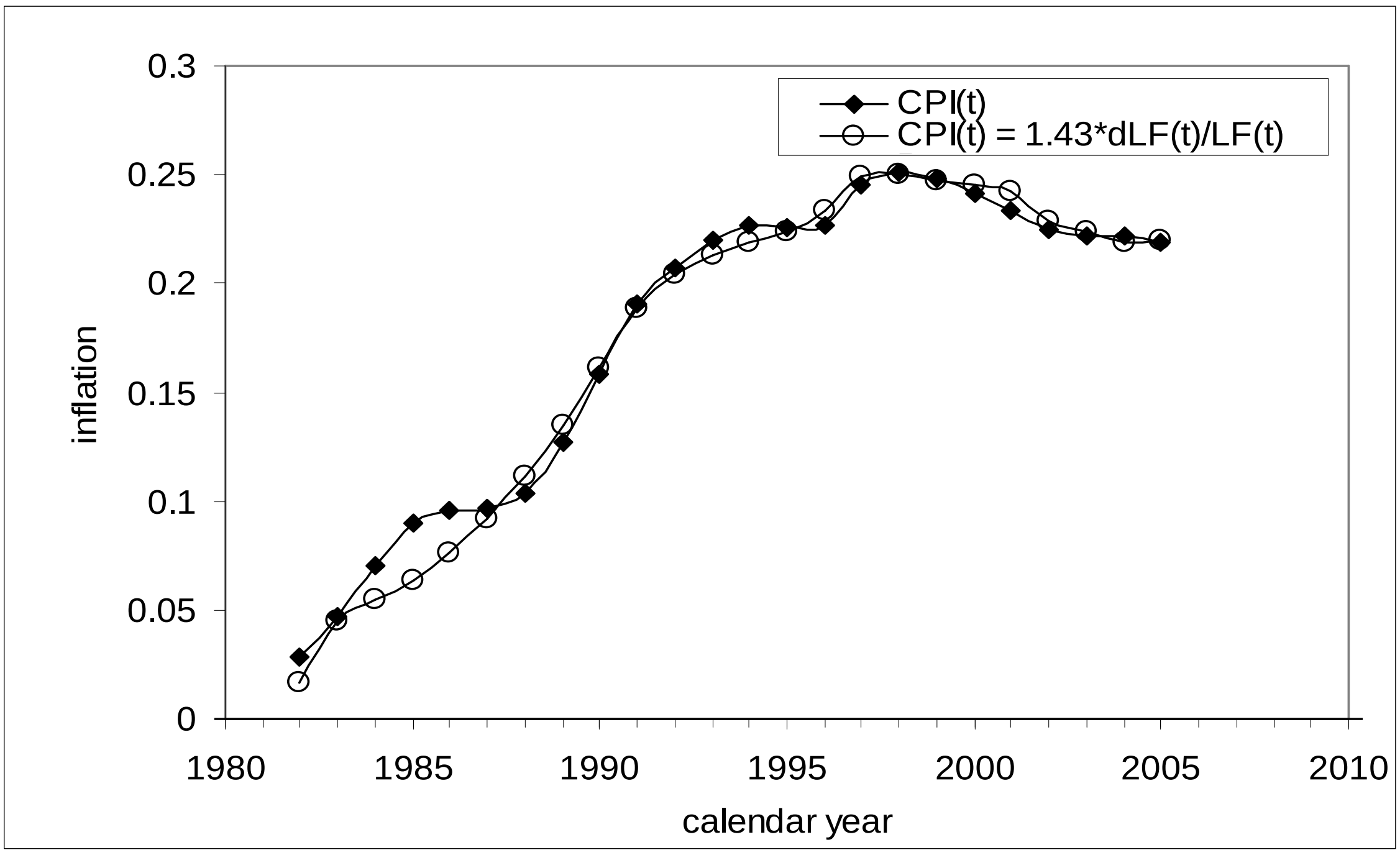

Figure 6. Measured inflation (CPI) and that predicted from the labor force change rate: a) dynamic curves; b) cumulative curves. Linear relationship between the variables is given in the upper right corner. A good agreement between the curves illustrates the prediction accuracy.

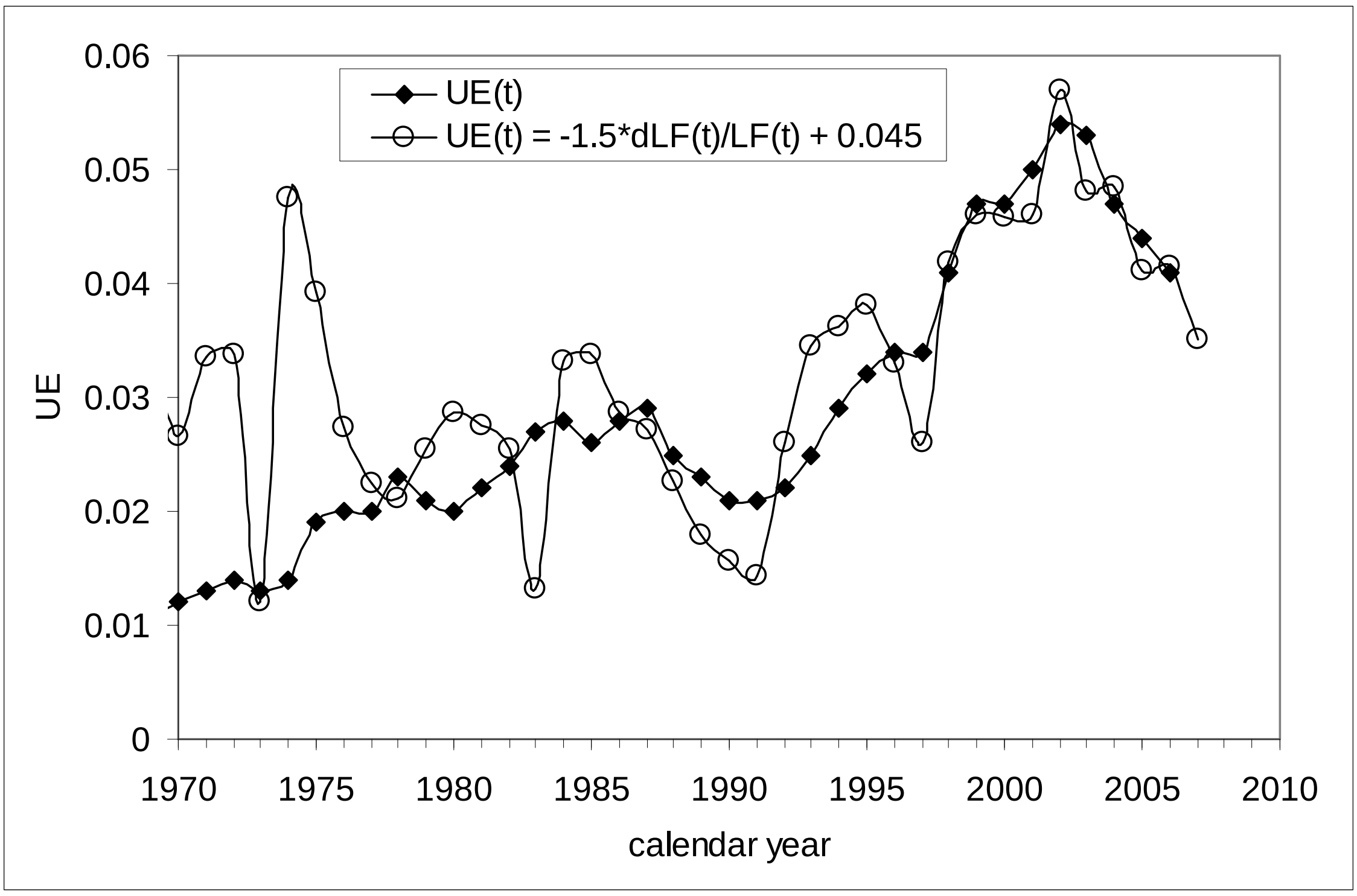

Figure 7. Comparison of measured unemployment and that predicted from the labor force change rate. Coefficients in the linear relationship are obtained by trail-and-error method as the best fit between the observed and predicted curve.

a)

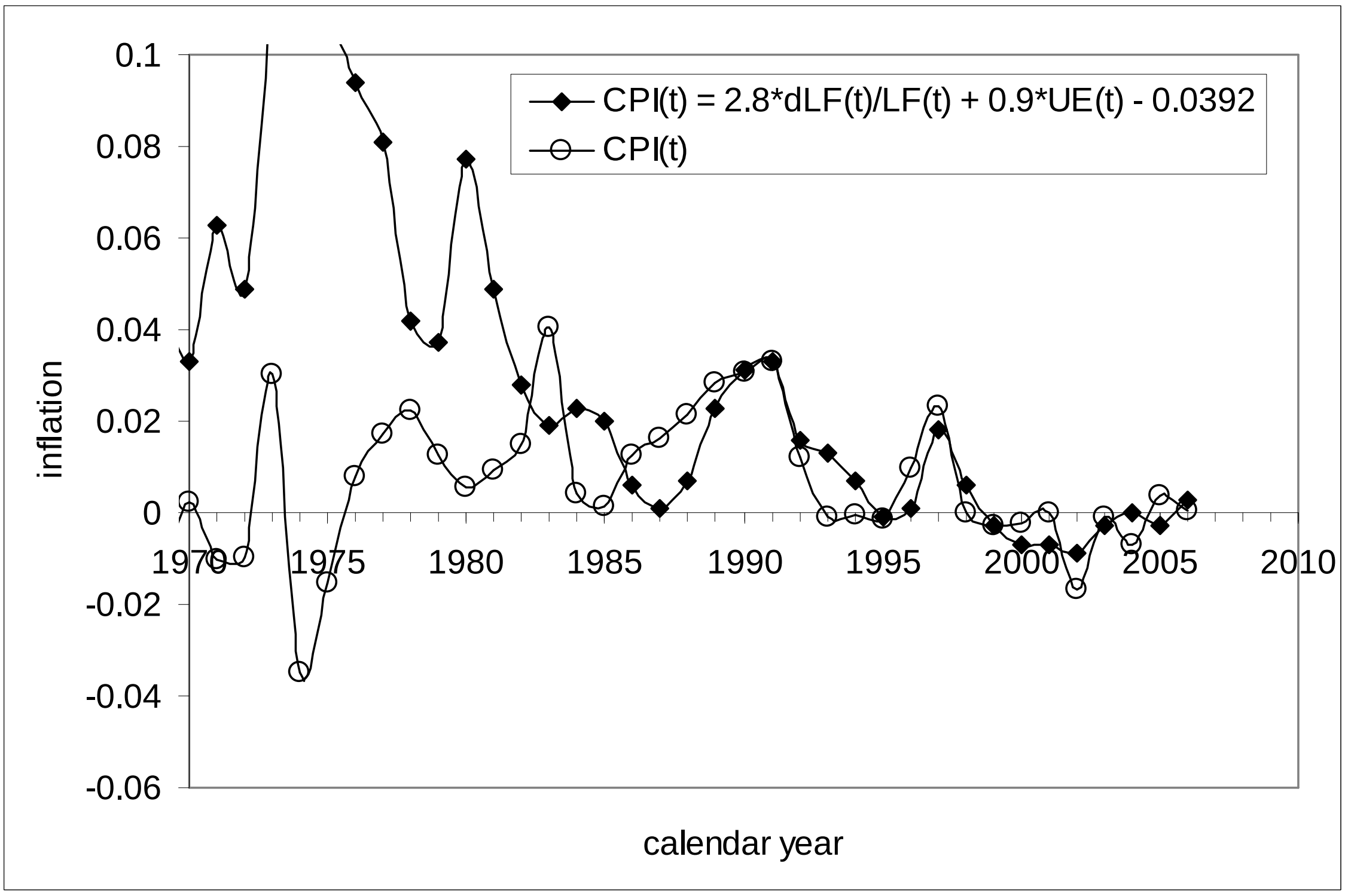

b)

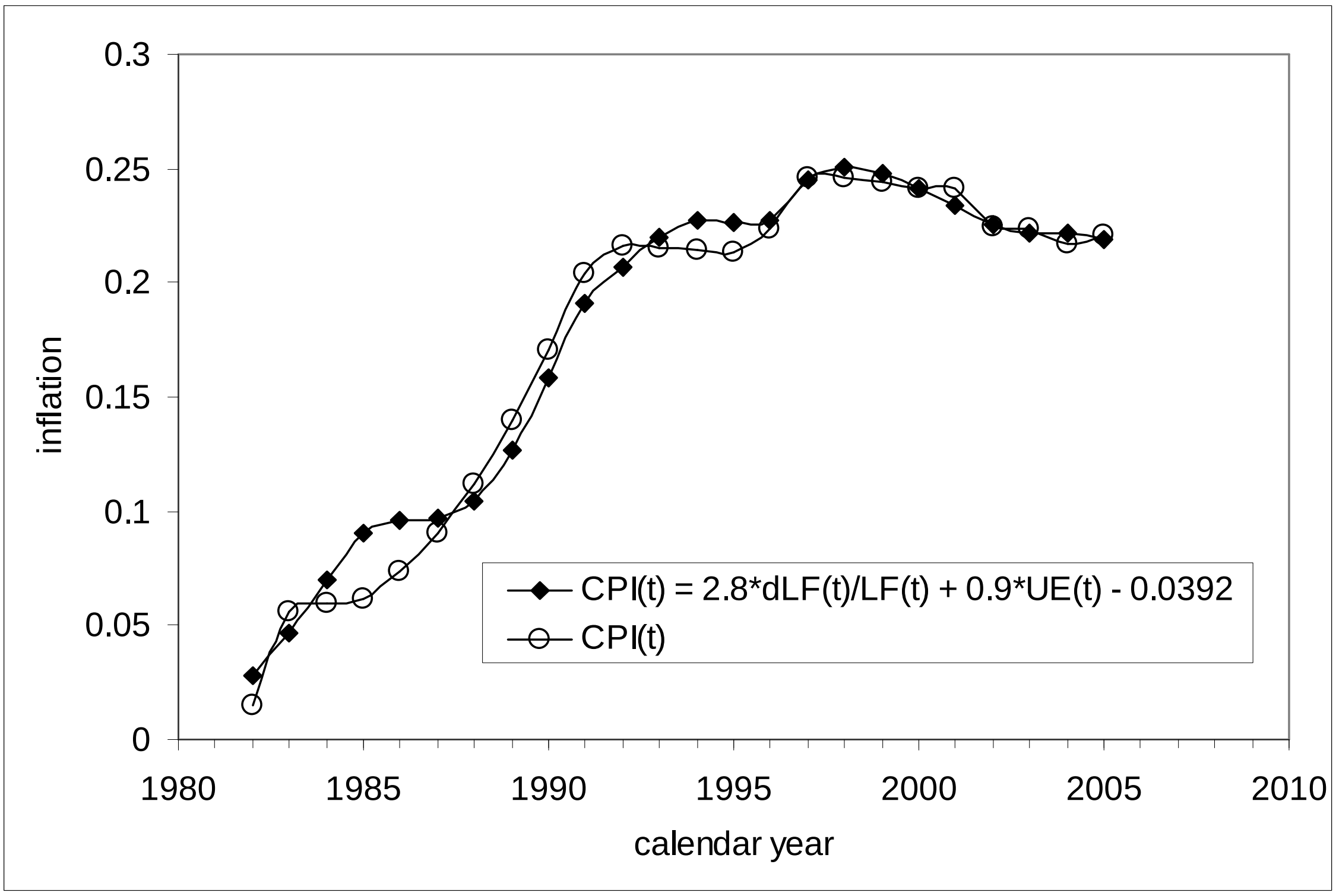

Figure 8. Comparison of the observed and predicted a) dynamic, and b**)** cumulative (starting from 1981) inflation curves.

Having relationships (2) and (3), one can easily predict the evolution of inflation and unemployment between 2007 and 2050 in Japan using various labor force projections. The National Institute of Population and Social Security Research (http://www.ipss.go.jp) provides quantitative projections of total population, which can be used for labor force projection. We consider the case of constant labor force participation rate fixed to 0.521 as measured in 2000. Figure 9 demonstrates the level of labor force in Japan will decrease from 67,000,000 in 2010 to 57,000,000 in 2050. Figure 10 displays the prediction of inflation and unemployment for the period through 2050. According to this prediction, 2007 is the last year of positive inflation (CPI) and Japan steps into a very long period of deflation.

**Conclusion**

There exists a Phillips curve for Japan with a negative coefficient of the linear link between inflation and unemployment, both variables evolving in sync. The existence of the Phillips curve does not facilitate the fight against deflation for the Bank of Japan. The deflationary period will last before the level of labor force will start to increase.

In Japan, the change rate of labor force level is the driving force behind unemployment and inflation. This finding confirms the existence of a generalized linear and lagged relationship between labor force, unemployment, and inflation in developed countries. The same relationship holds in the USA, France, Japan, Austria, Canada and Germany.

The change in labor force in Japan does no lead inflation and unemployment. This observation differs from those in other developed countries, where time lags as large as 6 years are observed (Germany). Labor force projections allow a reliable prediction of inflation and unemployment in Japan: CPI inflation will be negative (between -0.5% and -1% per year) in the next 40 years.  Unemployment will increase from 4.0% in 2010 to 5.3% in 2050

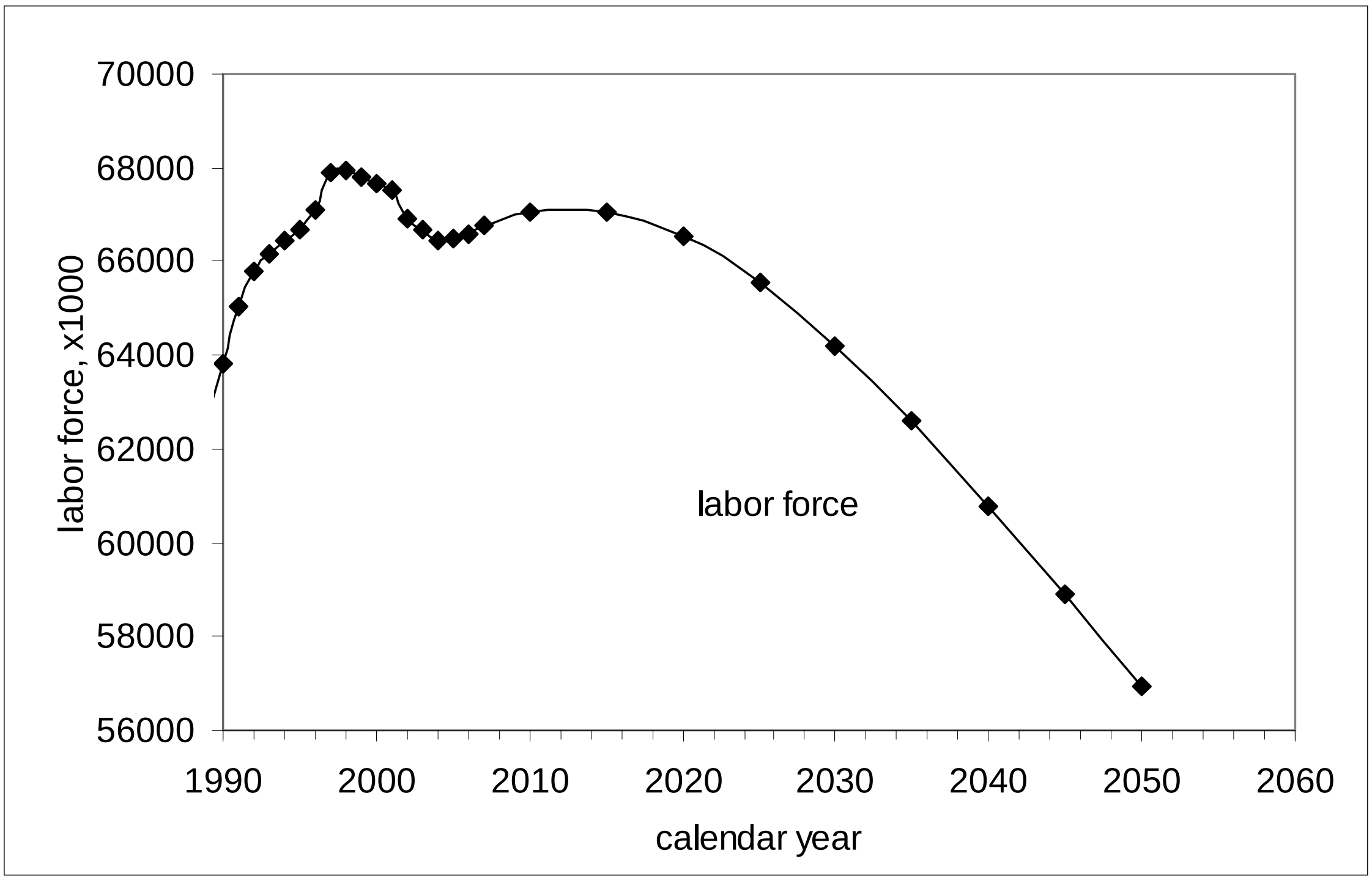

Figure 9. Projection of the labor force evolution between 2005 and 2050.

a)

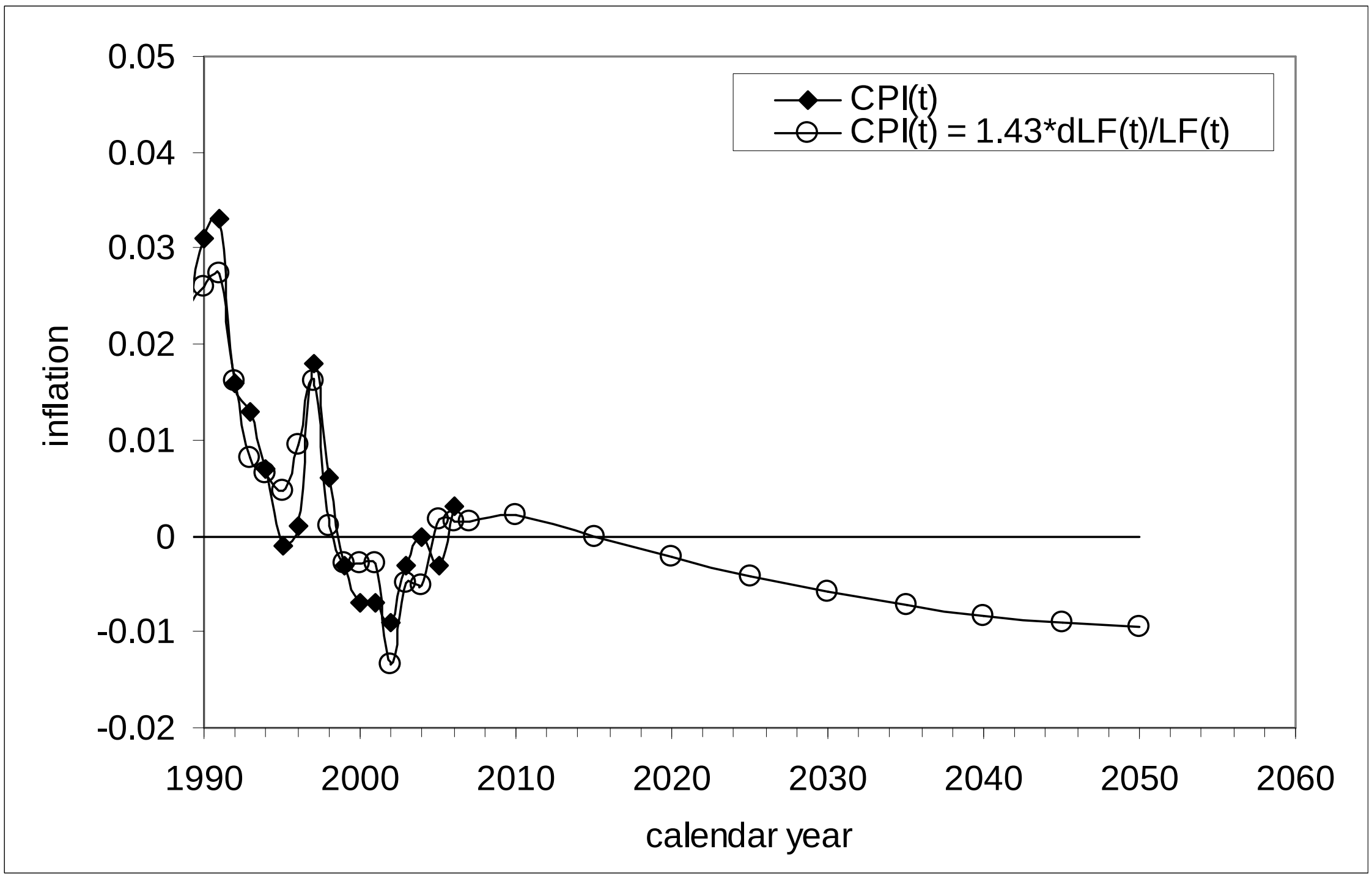

b)

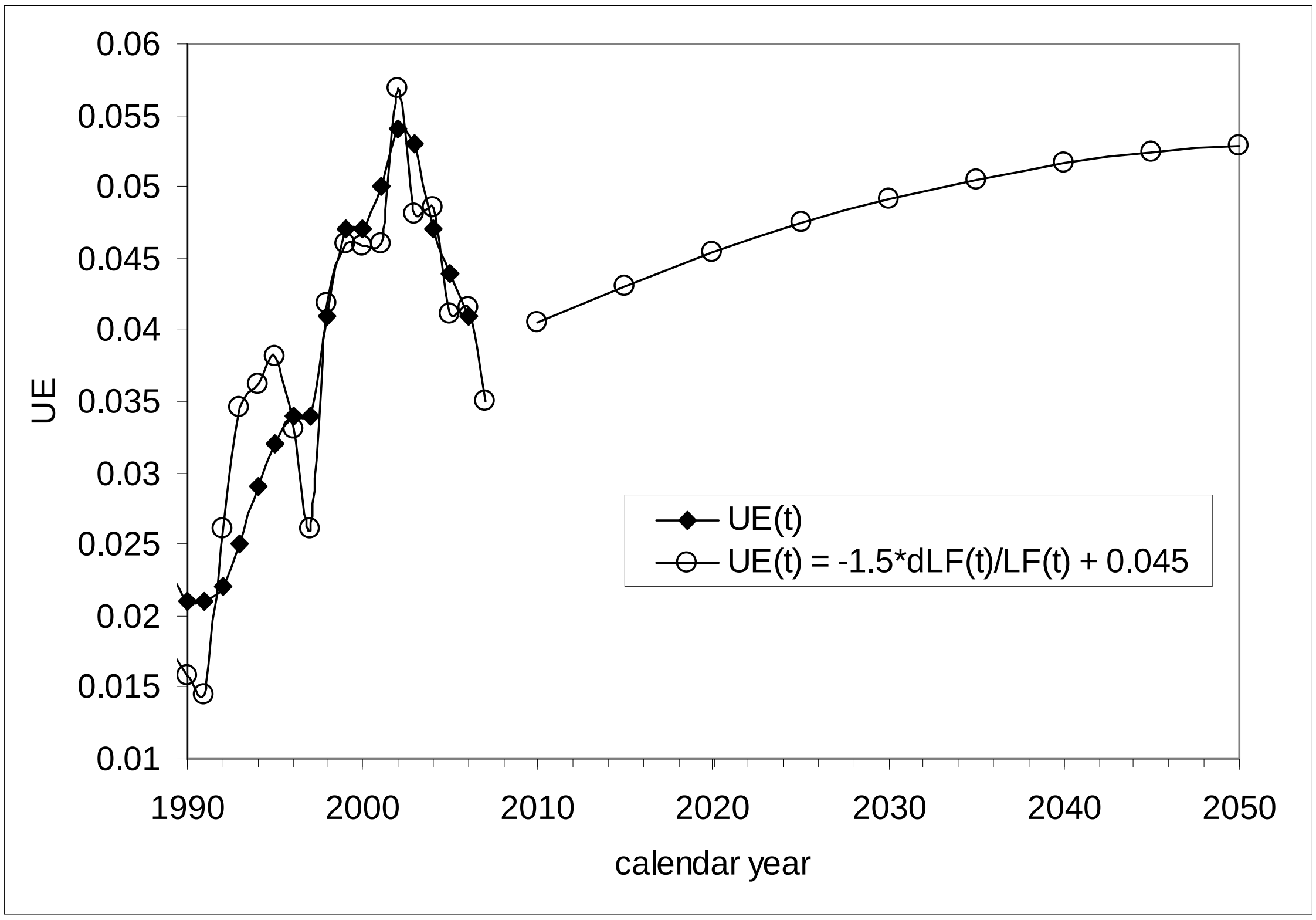

Figure 10. Prediction of the evolution of a) CPI inflation rate, and b) unemployment rate in Japan.

## References

Ariga, K., Matsui, K., (2002). Mismeasurement of CPI, http://www2.e.u-tokyo.ac.jp/~seido/output/Ariga/ariga029.pdf

Bureau of Labor Statistic, (2007). Foreign Labor Statistic. Table, retrieved at 20.07.2007 from http://data.bls.gov/PDQ/outside.jsp?survey=in

Caporale, G. M., Gil-Alana, L. A., (2006). Modeling structural breaks in the US, UK and Japanese unemployment rates, CESIFO Working Paper, 1734

De Veirman, E., (2007). Which Nonlinearity in the Phillips Curve? The Absence of Accelerating Deflation in Japan, Reserve Bank of New Zealand, January 14, 2007

Eurostat, (2006), User Queries, http://epp.eurostat.cec.eu.int/

Feyzioğlu, T., Willard, L., (2006). Does Inflation in China Affect the United States and Japan? IMF Working paper, WP/06/36.

Kamada, K., (2004). Real-Time Estimation of the Output Gap in Japan and its Usefulness for Inflation Forecasting and Policymaking, Deutsche Bank, Discussion Paper Series 1: Studies of the Economic Research Centre No 14/2004

Kitov, I., (2007a). Exact prediction of inflation and unemployment in Canada, MPRA Paper 5015, University Library of Munich, Germany.

Kitov, I., (2007b). Exact prediction of inflation and unemployment in Germany, MPRA Paper 5088, University Library of Munich, Germany.

Kitov, I., Kitov, O., Dolinskaya, S., (2007a). Relationship between inflation, unemployment and labor force change rate in France: cointegration test, MPRA Paper 2736, University Library of Munich, Germany

Kitov, I., Kitov, O., Dolinskaya, S., (2007b). Inflation as a function of labor force change rate: cointegration test for the USA, MPRA Paper 2734, University Library of Munich, Germany.

Kitov, I. (2006a). Inflation, unemployment, labor force change in the USA, Working Papers 28, ECINEQ, Society for the Study of Economic Inequality.

Kitov, I., (2006b). Exact prediction of inflation in the USA, MPRA Paper 2735, University Library of Munich, Germany.

Kitov, I., (2006c). The Japanese economy, MPRA Paper 2737, University Library of Munich, Germany.

Kitov, I., (2006d). Real GDP per capita in developed countries, MPRA Paper 2738, University Library of Munich, Germany.

Leigh, D., (2004). Monetary Policy and the Dangers of Deflation: Lessons from Japan Department of Economics, Johns Hopkins University, August 10, 2004

Nelson, E., (2006). The Great Inflation and Early Disinflation in Japan and Germany, Federal Reserve Bank of St. Louis, Working Paper 2006-052D, St. Louis, http://research.stlouisfed.org/wp/2006/2006-052.pdf

Organization for Economic Cooperation and Development, (2005). Notes by Country: Japan. http://www.oecd.org/dataoecd/35/3/ 2771299.pdf

Organization for Economic Cooperation and Development (2006), Corporate Data Environment, Labor Market Statistics, DATA, User Queries, January 8, 2006, http://www1.oecd.org/scripts/cde/

Pascalau, R., (2007). Productivity shocks, unemployment persistence, and the adjustment of real wages in OECD countries, available at: http://ssrn.com/abstract=962245

Rudd, J., Whelan, K., (2005). New tests of the New Keynesian Phillips curve, Journal of Monetary Economics, vol. 52, pages 1167-1181

Sekine, T., (2001). Modeling and Forecasting Inflation in Japan, IMF Working Paper, WP/01/82

Shiratsuka, S., (1999). Measurement errors in Japanese Consumer Price Index, Working Paper Series WP-99-2, Federal Reserve Bank of Chicago, http://ideas.repec.org/a/ime/imemes/v17y1999i3p69-102.html

The Statistics Bureau, the Ministry of Internal Affairs and Communications (2006), http://www.stat.go.jp/english/index.htm